\documentclass[12pt]{iopart}

\usepackage{amsbsy}
\usepackage{amssymb}
\usepackage{graphicx}
\usepackage{color}
\newcommand{\bea}{\begin{eqnarray}}
\newcommand{\eea}{\end{eqnarray}}
\newcommand{\beann}{\begin{eqnarray*}}
\newcommand{\eeann}{\end{eqnarray*}}
\newcommand{\be}{\begin{equation}}
\newcommand{\ee}{\end{equation}}

\begin{document}

\title[Correlations of heat and charge currents in quantum-dot thermoelectric engines]{Correlations of heat and charge currents in quantum-dot thermoelectric engines}

\author{Rafael S\'anchez}
\address{Instituto de Ciencia de Materiales de Madrid (ICMM-CSIC), Cantoblanco 28049, Madrid, Spain}
\author{Bj\"orn Sothmann}
\address{D\'epartement de Physique Th\'eorique, Universit\'e de Gen\`eve, CH-1211 Gen\`eve 4, Switzerland}
\author{Andrew N. Jordan}
\address{Department of Physics and Astronomy, University of Rochester, Rochester, New York 14627, USA}
\author{Markus B\"uttiker}
\address{D\'epartement de Physique Th\'eorique, Universit\'e de Gen\`eve, CH-1211 Gen\`eve 4, Switzerland}

\begin{abstract}
We analyze the noise properties of both electric charge and heat currents as well as their correlations in a quantum-dot based thermoelectric engine. The engine is a three-terminal conductor with crossed heat and charge flows where heat fluctuations can be monitored by a charge detector. We investigate the mutual influence of charge and heat dynamics and how it is manifested in the current and noise properties.
In the presence of energy-dependent tunneling, operating conditions are discussed where a charge current can be generated by heat conversion. In addition, heat can be pumped into the hot source by driving a charge current in the coupled conductor. An optimal configuration is found for structures in which the energy dependence of tunneling maximizes asymmetric transmission with maximal charge-heat cross-correlations. Remarkably, at a voltage that stalls the heat engine we find that in the optimal case the non-equilibrium state is maintained by fluctuations in the heat and charge currents only.
\end{abstract}

\maketitle

\section{Introduction}


Thermoelectric effects have been a subject in the physics of mesoscopic conductors almost since the beginning of the field, both experimentally~\cite{molenkamp,molenkamp2} and theoretically~\cite{sivan,butcher,cb}. Only in recent years has the way to manipulate electronic heat currents at the nanoscale found a broader interest for its potential applications~\cite{giazotto} and non-linear properties~\cite{david,meair,whitney,dutt}. Among the different nanoscale devices, quantum dots with discrete energy levels have stood out because of their use as tunable energy filters~\cite{federica}. Based on that effect, two-terminal quantum-dot based refrigerators~\cite{edwards,smith}, heat pumps~\cite{misha,liliana,miguel,juergens}, heat diodes~\cite{ruokola} and heat converters~\cite{humphrey,muralidharan} have been proposed which are predicted to attain a high efficiency. 

In order to address the demanding and important issue of harvesting waste heat to generate useful power it is advantageous to have a three-terminal setup. Two terminals at the same temperature support a charge current with the third (hot) one providing only heat. 
Having charge and heat flowing in different directions, three terminal geometries profit from allowing the thermal and electric forces to be applied to different terminals. Also the charge and heat currents as well as their correlations can be measured separately. 
Recent proposals include models for heat transfer mediated by electron-electron interaction~\cite{hotspots,detection,cavities,venturelli}, electron-boson coupling~\cite{entin,jiang,bjorn}, an electromagnetic environment~\cite{ruokola2}, or a generic heat source~\cite{saito,david2,cheese,wells}. In such geometries, a finite electronic current is generated between the two terminals of an unbiased conductor by correlating the transfer of charge and the absorption of energy from the hot source. The effect  can be quantified by writing a generalized fluctuation-dissipation theorem that relates the equilibrium cross-correlations, $S_{IJ}$, of charge, $I$, and absorbed heat currents, $J$, to ``crossed'' Seebeck and Peltier coefficients:
\be
\label{fdt}
S_{IJ}^{eq}=2kT\left.\frac{\partial I}{\partial (\Delta T/T)}\right|_{\Delta V=0}=2kT\left.\frac{\partial J}{\partial \Delta V}\right|_{\Delta T=0}.
\ee
Note that here the two currents respond to thermodynamic forces applied to different conductors: the charge current $I$ responds to the increased temperature in the third terminal ($\Delta T=T_3-T$), while the heat current through the hot contact, $J$, responds to the voltage applied to the conductor ($\Delta V=V_1-V_2$). 
The required correlations appear in the presence of an appropriate symmetry breaking. The first equality in (\ref{fdt}) demonstrates their relevance for thermoelectric conversion. Configurations of interacting multiterminal conductors have been investigated with a view to enhanced correlations~\cite{mcclure,eugene2007,marlies,haupt,michalek,robert} and noise-induced transport~\cite{khrapai,levchenko,drag,laroche,news}, relevant in the context of fluctuation theorems~\cite{drag,utsumi,kung,bulnes,krause,detection}.

\begin{figure}[t]
\begin{center}
\includegraphics[width=0.8\linewidth,clip]{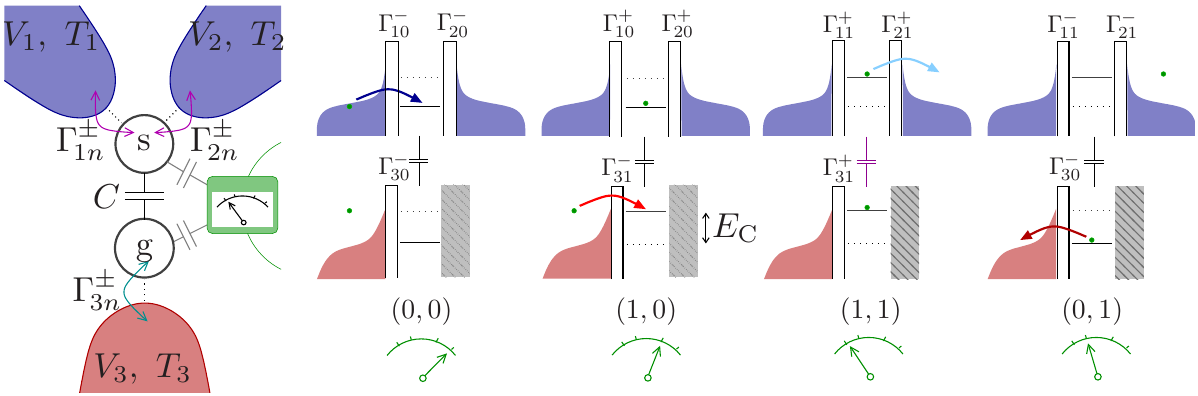}
\end{center}
\caption{\label{sys} A single electron energy harvester. A quantum dot ($s$) is coupled to two terminals at temperature $T_1=T_2=T$ and supports charge transport. The other one ($g$) is coupled to a third reservoir at a higher temperature, $T_3$, which will work as a heat source. A voltage $V_l$ can be applied to every terminal, and $\Gamma_{ln}^\pm$ are the tunneling rates through terminal $l$ with a charge $n$ in the other dot. Energy is transferred between the two dots by electron-electron interaction, parametrized by the capacitive coupling, $C$. A charge detector is asymmetrically coupled to both dots so it is able to detect the occupation of the four charge states of the coupled system, $(n_d,n_g)$, with $n_\alpha=0,1$. They are represented in the right panel forming a sequence that transfers an energy $E_C$ from the hot to the cold conductor.}
\end{figure}

Here, we investigate the charge- and heat-current auto- and cross-correlations in a three-terminal quantum-dot energy harvester schematically depicted in Figure \ref{sys}. A quantum dot is tunnel-coupled to two leads between which a charge current can flow. In addition, it is capacitively coupled to a second quantum dot such that they exchange energy via Coulomb interaction but no charge~\cite{bridge,hubel}. The charge of the second dot fluctuates due to the contact to a third terminal. When set at a higher temperature, this third fermionic reservoir will constitute our heat source. We consider the case when the heat flow from the hot source into the charge conductor is mediated solely by the electron-electron interaction. In a recent work, we have proposed a way to measure the statistics of heat transfer fluctuations in this model~\cite{detection}. The charge and heat transfer events can be resolved in time by a nearby charge detector, e.g. a quantum point contact. If the detector is asymmetrically coupled to the two dots, it can distinguish states with a different charge distribution in each dot~\cite{fujisawa,kung} and track the sequences that involve the transfer of energy between the conductors. In this way, the heat transfer statistics and its effect on the electronic transport can be extracted by standard electron-counting techniques. Recent experiments have measured the heat and entropy production statistics in a single-terminal quantum dot~\cite{saira,koski}.

In the Coulomb-blockade regime, where electron-electron interaction is strong, the charge of the system is well described by four states $(n_s, n_g)$ where $n_s$ is the occupation of the cold dot $s$, and $n_g$ that of the hot dot $g$, with $n_\alpha\in\{0,1\}$.
Each charge state will leave a distinct signal in the detector. For example, a series of single-electron transitions like $(0,0)\rightarrow(1,0)\rightarrow(1,1)\rightarrow(0,1)\rightarrow(0,0)$, cf. Figure \ref{sys}, transfers a fixed energy $E_C$ (determined by geometric capacitances)~\cite{hotspots} into the conductor $s$ causing an electron to tunnel from one lead to the other. Such state-resolved counting determines the number of electrons transferred to the hot terminal conditioned on the charge occupation of the conduction dot ($n=0,1$), $N_{gn}$. The energy deposited in the hot terminal by electrons tunneling from each state is $E_{gn}$, with $E_{g1}-E_{g0}=E_C$. Thus, after an electron has tunneled in and out of dot $g$, the energy that has been dissipated in the process can only be 0 or $\pm E_C$. This allows us to write the heat current in terms of particle counting~\cite{detection}:
\be
\label{strescount}
J=\frac{E_C}{2}\langle \dot N_{g1}-\dot N_{g0}\rangle,
\ee
as well as higher order cumulants\footnote{Note a factor 2 corrected which was not in the definition given in Ref.~\cite{detection}}. Note that the amount of transferred heat does not rely on the energy of tunneling electrons (as is the case for the heat current flowing through contacts $l=1,2$, which is proportional to $E_{sn}-qV_l$) so its detection is therefore not affected by voltage fluctuations. 

Usually, heat is carried by bosonic excitations such as phonons. Their fluctuations are thus expected to be super-Poissonian, i.e., their variance is larger than their mean. In contrast, noninteracting electron statistics is sub-Poissonian due to the Pauli exclusion principle~\cite{markus}.
When the energy dissipation and the fermionic currents become correlated, the character of their fluctuations can be mutually affected~\cite{rf}. 
We find configurations where their cross-correlations are maximal. In that case, the heat and charge current statistics become identical. Applied to thermoelectrics, such configurations are useful for achieving a tight energy-matter coupling, $I\propto J$, which is required for Carnot efficient converters in two-~\cite{humphrey,esposito} and three-terminal~\cite{hotspots} setups. Notably in our model, heat transfer is mediated solely by Coulomb interaction and therefore its statistics will strongly depend on electronic transport parameters. The fluctuations of heat carried by electrons have been investigated in two terminal conductors~\cite{averin,sergi,zhan} and single particle sources~\cite{battista}.

\section{Model and technique}
\label{sec:model}

We consider single-level quantum dots with chemical potentials $E_{\alpha,n}=\varepsilon_\alpha+U_{\alpha,n}$, where the index $\alpha=\{s,g\}$ refers to the charge conducting ($s$) and the hot gate quantum dot ($g$), with $n=\{0,1\}$ being the charge occupation of the other dot. It includes the bare energy of the level, $\varepsilon_\alpha$, and the charging energy, $U_{\alpha,n}$, which depends on the applied voltages, $V_l$, and must be calculated self-consistently. This is done by considering the capacitances associated to every tunnel barrier, $C_i$ and the interdot capacitance $C$, as shown in the Appendix. 
Remarkably, $U_{\alpha,1}=U_{\alpha,0}+E_C$, such that electrons need to have an extra energy $E_C=2q^2/\tilde C$ (fixed by the elementary charge $q$ and the effective capacitance $\tilde C$) in order to tunnel into a quantum dot when the other one is already occupied. An energy amount $E_C$ is transferred between the two conductors if the consecutive events of an electron tunneling in and out of a quantum dot occur at different occupations of the coupled system. One such sequence is illustrated in figure \ref{sys} which correlates the transfer of an energy $E_C$ with a charge $q$ being transported between different terminals of the conductor.
Hence, $E_C$ plays the role of the quantum of transferred heat as $q$ is the quantum of transferred charge~\cite{hotspots}, see  (\ref{strescount}) or (\ref{zbcharge}) and (\ref{zbheat}) below. This process gives a maximal cross-correlation contribution.

\subsection{Counting statistics}

We consider the situation where the dots are weakly coupled to the leads such that transport occurs via sequential tunneling. The electron dynamics is well described by a Markovian master equation for the reduced density matrix~\cite{cb} with diagonal elements given by the occupation probability of each charge state, $\rho=(\rho_{00},\rho_{10},\rho_{01},\rho_{11})^T$. 

We are interested in the charge flowing through one of the leads of the conductor (say $l{=}2$) and the heat transferred from the heat source (contact $l{=}3$). The stationary charge and heat currents, $I{=}\langle\hat I\rangle$ and $J{=}\langle\hat J\rangle$, zero frequency charge and heat noise, $S_{II}{=}\langle(I-\hat I)^2\rangle$ and $S_{JJ}{=}\langle(J-\hat J)^2\rangle$, and cross-correlation, $S_{IJ}{=}\langle(I-\hat I)(J-\hat J)\rangle$, as well as higher order cumulants, can be calculated by counting statistics techniques~\cite{bagrets,andrieux2}\footnote{The formalism can be extended to the combined counting of several charge, energy or spin currents by considering different counting fields~\cite{eugene2007,rf,christian,sb,braggio}.}.
Here we introduce the charge, $\chi$, and heat, $\xi$, counting fields\footnote{We recall that our system allows for the measurement of heat transfer statistics in terms of state resolved counting. One can thus equivalently obtain the same information by replacing $e^{\pm i\Theta_{3n}\xi}$ in (\ref{MLR}) by $e^{\pm i\xi_n}$, with state-resolved {\it particle} counting fields $\xi_n$ that account for the number of tunneled electrons $N_{gn}$. However, one has to deal then with three different counting fields so, for theoretical simplicity it is easier to consider a single {\it heat} counting field, $\xi$. Furthermore, in order to detect the charge current statistics in the conductor, a more complicated structure allowing for a directional counting --such as a double quantum dot-- would be needed~\cite{fujisawa,kung}.} in the master equation, $\dot\rho={\cal M}(\chi,\xi)\rho$, with~\cite{detection}:
\begin{eqnarray}\label{MLR}
\fl
{\cal M}(\chi,\xi)=
\left(\begin{array}{cccc}
-\Gamma_{\!s0}^-{-}\Gamma_{\!g0}^- &\Gamma_{\!10}^+{+}e^{iq\chi}\Gamma_{\!20}^+& e^{i\Theta_{30}\xi}\Gamma_{\!g0}^+ & 0\\
\Gamma_{\!10}^-{+}e^{-iq\chi}\Gamma_{\!20}^- & \ \ -\Gamma_{\!s0}^+{-}\Gamma_{\!g1}^- & 0 & e^{i\Theta_{31}\xi}\Gamma_{\!g1}^+\\
e^{-i\Theta_{30}\xi}\Gamma_{\!g0}^- & 0 & -\Gamma_{\!s1}^-{-}\Gamma_{\!g0}^+ & \Gamma_{\!11}^+{+}e^{iq\chi}\Gamma_{\!21}^+\\
0 & e^{-i\Theta_{31}\xi}\Gamma_{\!g1}^- & \Gamma_{\!11}^-{+}e^{-iq\chi}\Gamma_{\!21}^- & -\Gamma_{\!s1}^+{-}\Gamma_{\!g1}^+\\
\end{array}  \right),
\end{eqnarray}
where $\Theta_{ln}=E_{\alpha,n}-qV_l$ is the heat carried by electrons in the tunneling process, and $\Gamma_{\!sn}^\pm{=}\Gamma_{\!1n}^\pm{+}\Gamma_{\!2n}^\pm$. Here, $\Gamma_{ln}^\pm$ are the tunneling rates for electrons tunneling into ($-$) or out of ($+$) a quantum dot through barrier $l$ when the other dot contains $n$ electrons. 
They are given by Fermi's golden rule as $\Gamma_{ln}^-=\Gamma_{ln}f(\Theta_{ln}\beta_l)$ and $\Gamma_{ln}^+=\Gamma_{ln}-\Gamma_{ln}^-$, with the Fermi function $f(x)=(1+e^x)^{-1}$ and $\beta_l=(kT_l)^{-1}$. Note that the energy dependence of the  rates arises both from the Fermi functions and the tunnel couplings, $\Gamma_{ln}$. We consider the regime $kT_l\gg\Gamma_{ln}$ where broadening of the energy levels can be neglected. Currents are defined as positive when flowing into terminals.

The correlations are given by derivatives of the cumulant generating function, ${\cal F}(\chi,\xi)$, with respect to the counting fields. ${\cal F}$ is the eigenvalue of ${\cal M}(\chi,\xi)$ that approaches 0 as  $\chi,\xi\to0$. In general, an analytical expression for $\cal F$ cannot be obtained. In this case, the cumulants can still be extracted order by order. Following Ref.~\cite{rf}, we make an expansion
${\cal F}(\chi,\xi)=\sum_{m,p>0}c_{m,p}(e^{i\chi}-1)^m(e^{i\xi}-1)^p$
such that we get recursively the stationary currents, $I=c_{1,0}$, $J=c_{0,1}$, zero frequency noises, $S_{II}=I+2c_{2,0}$ and $S_{JJ}=J+2c_{0,2}$, and cross-correlations, $S_{IJ}=c_{1,1}$. 

It is useful to normalize the auto-correlations such that the charge and heat Fano factors, $F_I=S_{II}/qI$ and $F_J=S_{JJ}/E_CJ$, reflect the sub- or super-Poissonian character of the fluctuations when they are lesser or greater than 1, respectively. We can also define the cross-correlation coefficient, $r=S_{IJ}/\sqrt{S_{II}S_{JJ}}$, which is bounded by the Cauchy-Schwartz inequality: $-1\le r\le 1$. We have $r=\pm1$ when the two currents are totally correlated such that their cumulants are proportional to each other.


\subsection{Stability diagram}

\begin{figure}[t]
\begin{center}
\includegraphics[width=0.7\linewidth,clip]{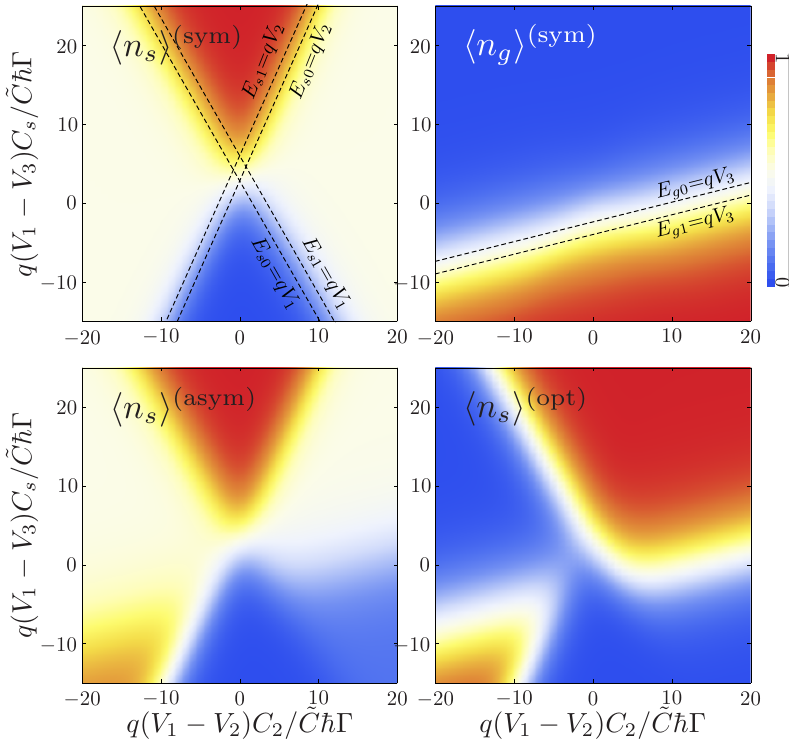}
\end{center}
\caption{\label{occup} Occupation of each dot, $\langle n_\alpha\rangle$, as function of the applied voltages for different tunneling rate configurations. Charge fluctuations occur in the regions $0<\langle n_\alpha\rangle<1$. In the symmetric configuration, all tunnel couplings are equal and energy independent, $\Gamma_{ln}=\Gamma$. For the asymmetric one, we have $\Gamma_{11}=0.1\Gamma$. The optimal configuration is given by $\Gamma_{10}=\Gamma_{21}=\Gamma$ and $\Gamma_{11}=\Gamma_{20}=0$. In the plots of $\langle n_s\rangle^{(sym)}$ and $\langle n_g\rangle^{(sym)}$, the alignment of the four chemical potentials $E_{\alpha n}$ with those of the different leads is represented by dashed lines. In the symmetric configuration the occupation $\langle n_s\rangle^{(sym)}$ shows a typical Coulomb blockade stability diagram. The blue and red plateaux in  $\langle n_s\rangle^{(sym)}$ correspond to the Coulomb-blockade region where dot $s$ is either empty or occupied by one electron. Out of them, charge fluctuates and current can flow. $\langle n_g\rangle^{(sym)}$ fluctuates along the lines $E_{gn}=qV_3$. In the presence of energy-dependent tunnel couplings, $\langle n_s\rangle^{(asym)}$ and $\langle n_s\rangle^{(opt)}$ are affected by the occupation of dot $g$.  Parameters: $kT_1{=}kT_2{=}kT_3/2{=}5\hbar\Gamma$, $q^2/C_i{=}20\hbar\Gamma$, $q^2/C{=}50\hbar\Gamma$, $\varepsilon_{s}{=}\varepsilon_{g}=0$.}
\end{figure}

From the stationary solution of the master equation ${\cal M}(0,0)\bar\rho=0$, we obtain the occupation probabilities of the different charge states. The analytical expressions are given in Ref.~\cite{hotspots}. Using them, one gets the dot occupations
$\langle n_s\rangle=\bar\rho_{10}+\bar\rho_{11}$ and $\langle n_g\rangle=\bar\rho_{01}+\bar\rho_{11}$.
Their dependence on the two voltage differences $V_1-V_2$ (the applied bias) and $V_1-V_3$ (which serves as a gate voltage to dot $s$) defines the stability diagram of the system. They are also affected by the coupling to the leads, in particular by their asymmetry due to energy-dependent tunneling. We define here three different configurations that will be considered in the following:
\begin{itemize}
\item Symmetric configuration: $\Gamma_{ln}=\Gamma$ for all  $l,n$;
\item Asymmetric configuration: $\Gamma_{10}\Gamma_{21}\ne\Gamma_{11}\Gamma_{20}$, where noise-induced transport takes place. For our numerics, we choose it to be as in the symmetric configuration except that $\Gamma_{11}<\Gamma$;
\item Optimal configuration: transitions in the conductor $(0,n_g)\leftrightarrow(1,n_g)$ with different $n_g=0,1$ involve different terminals. Here, we will consider couplings as in the symmetric case except for $\Gamma_{11}=\Gamma_{20}=0$.
\end{itemize}
In the region delimited by the conditions $E_{s1}=qV_1$ and $E_{s1}=qV_2$, the occupation of dot $s$ is fixed (0 or 1) by Coulomb blockade, see figure \ref{occup}. For larger bias voltage, the charge fluctuates by tunneling and shot noise is present. Being coupled to only one terminal, the charge of dot $g$ only fluctuates along $E_{gn}=qV_3$. We call it the {\it fluctuating gate} region whose width is $kT_3$. In the presence of energy-dependent tunnel couplings, the regions where $n_s$ fluctuates are modified by
the presence of the coupled dot, $g$, see $\langle n_s\rangle^{(asym)}$ and $\langle n_s\rangle^{(opt)}$ in figure \ref{occup}, which will affect the non-equilibrium currents. 

\section{Transport and correlations}

In what follows, we analyze the dynamics of charge and heat and their correlations for the different regimes defined by asymmetry and applied voltages.

\subsection{Zero bias transport}

For our system to work as an energy harvester, both left-right and particle-hole symmetry have to be broken. Otherwise at zero applied bias, transport is equiprobable in opposite directions and no net current is generated. For the quantum-dot harvester, the symmetry breaking is provided by asymmetric, energy-dependent tunnel couplings $\Gamma_{ln}$~\cite{hotspots,cavities,ruokola2}.
Then, processes carrying an electron from left to right or vice versa after absorbing energy from the hot bath will have different rates (proportional to $\Gamma_{10}\Gamma_{21}$ and  $\Gamma_{20}\Gamma_{11}$, respectively), resulting in a finite current generated at zero bias~\cite{hotspots}:
\be
\label{zbcharge}
I=q\frac{(\Gamma_{11}\Gamma_{20}-\Gamma_{10}\Gamma_{21})\Gamma_{g0}\Gamma_{g1}}{8\gamma^3}\sinh\left[\frac{E_C}{2}\left(\beta_3-\beta\right)\right]\prod_{\alpha,n}\mathrm{sech}\left(\frac{\Theta_{\alpha,n}\beta_\alpha}{2}\right),
\ee
if the two conductors are at different temperatures. Here, $\Theta_{\alpha,n}=E_{\alpha,n}-qV_\alpha$, with $V_1=V_2=V_s$ and $V_g=V_3$. Through all the paper, we will consider $\beta_3<\beta_1=\beta_2=\beta$.
The denominator $\gamma^3$ contains third order products of tunnel couplings. In contrast to the charge current, heat flows into the conductor independently of the asymmetry: 
\be
\label{zbheat}
J=E_C\frac{\Gamma_{s0}\Gamma_{s1}\Gamma_{g0}\Gamma_{g1}}{8\gamma^3}\sinh\left[\frac{E_C}{2}\left(\beta_3-\beta\right)\right]\prod_{\alpha,n}\mathrm{sech}\left(\frac{\Theta_{\alpha,n}\beta_\alpha}{2}\right).
\ee
The finite charge and heat current cross-correlations are evident already in the proportionality of the two currents (\ref{zbcharge}) and (\ref{zbheat}). The explicit expression for the equilibrium cross-correlations, as discussed in (\ref{fdt}), is:
\be
\label{eqcrosscorr}
S_{IJ}^{eq}=2qE_C\frac{(\Gamma_{11}\Gamma_{20}-\Gamma_{10}\Gamma_{21})\Gamma_{g0}\Gamma_{g1}}{8\gamma^3}\prod_{\alpha,n}\mathrm{sech}\left(\frac{\Theta_{\alpha,n}\beta_\alpha}{2}\right),
\ee
which clearly shows that the asymmetry $\Gamma_{11}\Gamma_{20}-\Gamma_{10}\Gamma_{21}$ is necessary in order to correlate the heat and charge dynamics in the linear regime. From (\ref{zbcharge}) and (\ref{eqcrosscorr}) one can easily verify the first equality in the fluctuation-dissipation theorem for crossed charge and heat currents (\ref{fdt}), the second one being fulfilled by multiterminal Onsager symmetry~\cite{onsager,philippe,matthews,hwang}. For completeness, we have checked that the auto-correlations obey $S_{II}^{eq}{=}2kT\partial I/\partial\Delta V$ and $S_{JJ}^{eq}=2kT\partial J/\partial(\Delta T/T)$.

\begin{figure}[t]
\begin{center}
\includegraphics[width=0.4\linewidth,clip]{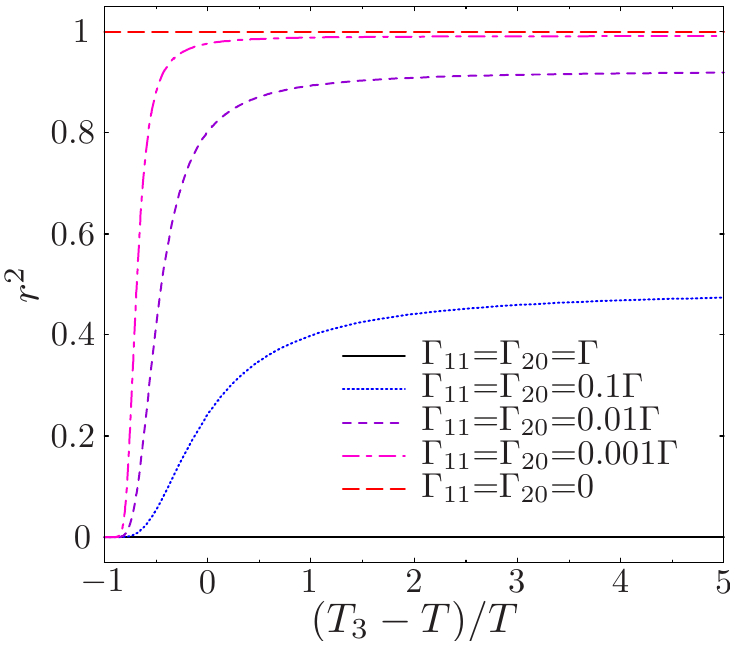}
\end{center}
\caption{\label{r2} Zero bias cross-correlation coefficient at zero bias ($V_1=V_2$) as a function of the temperature gradient for different tunneling asymmetries. All the tunnel couplings are considered equal, $\Gamma_{ln}{=}\Gamma$, except those explicitly mentioned in the legend. For the totally symmetric configuration ($\Gamma_{11}{=}\Gamma_{20}{=}\Gamma$), the heat and charge currents are uncorrelated and no charge current will be generated. In the optimal configuration, $\Gamma_{11}{=}\Gamma_{20}{=}0$, heat and charge flows are maximally correlated, with one electron being transferred between the two leads of the conductor for every energy $E_C$ absorbed from the hot source.}
\end{figure}

As shown in figure \ref{r2}, the cross-correlations for $V_1=V_2$ vanish in the perfectly symmetric case, $\Gamma_{ln}=\Gamma$ and increase with the asymmetry. In the particular case where each charge state of the conductor is coupled to only one lead (e.g. $\Gamma_{11}=\Gamma_{20}=0$), the cross-correlation is maximal, $r^2=1$, with an electron being transported across $s$ for every energy quantum $E_C$ absorbed from the hot source. We call it the optimal configuration for achieving a tight energy-matter coupling, $I^{(opt)}/q=J^{(opt)}/E_C$, for which the converter reaches the Carnot efficiency~\cite{hotspots} and the maximal harvesting thermopower $E_C/(qT_3)$~\cite{detection}.

\subsection{Charge and heat currents}

\begin{figure}[t]
\begin{center}
\includegraphics[width=\linewidth,clip]{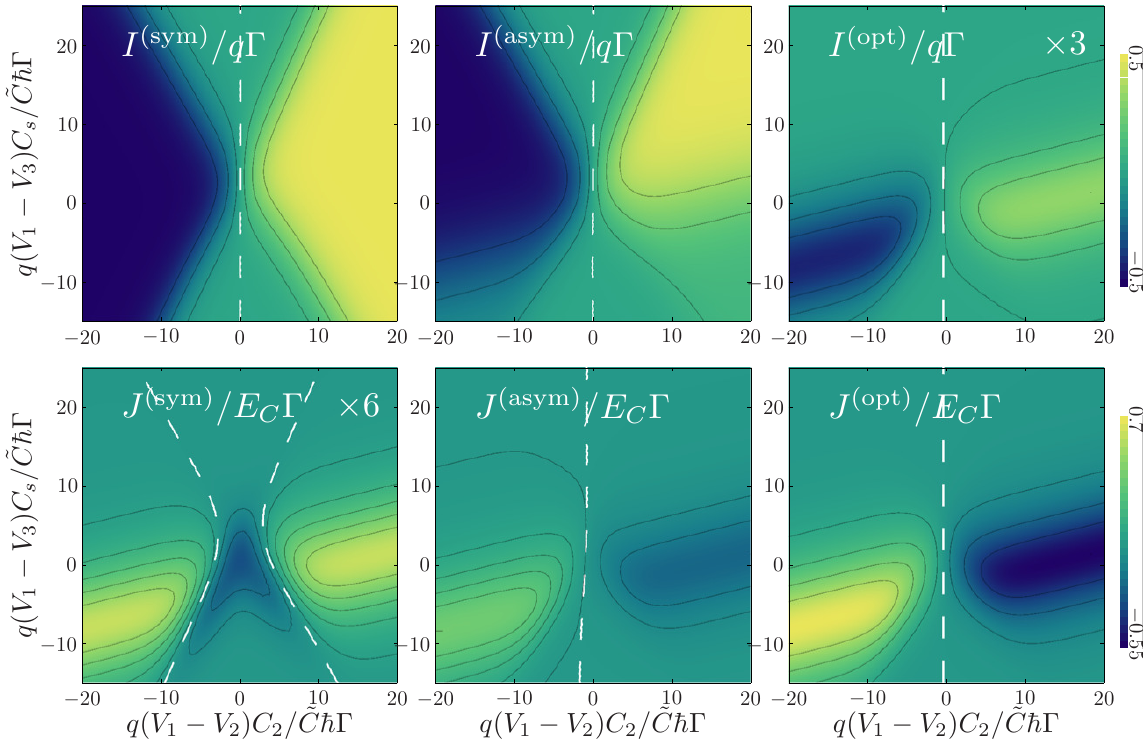}
\end{center}
\caption{\label{curr} Charge, $I$, and heat, $J$, currents as function of the applied voltages for the different tunneling rate configurations. 
In the symmetric configuration the charge current $I^{(sym)}$ shows a typical Coulomb blockade stability diagram. It is affected by the injected heat in the asymmetric configurations, $I^{(asym)}$ and $I^{(opt)}$. The heat current only flows for voltages around the conditions $\Theta_{3n}\equiv E_{3n}-qV_3=0$ (the fluctuating gate region) and can be reversed by voltage in the presence of asymmetry. White dashed lines indicate the points of vanishing current. Parameters as in figure \ref{occup}.}
\end{figure}

The nonlinear transport is governed by the alignment of the different states with the Fermi levels of the leads. They correspond to the conditions $\Theta_{ln}=E_{ln}-qV_l=0$ (marked by dotted lines in figure \ref{occup}), where charge fluctuations through the corresponding contact are enhanced. Far from them, the charge of at least one of the dots is well defined and the currents show plateaux as a function of the applied voltages. In our three-terminal system, not only the voltage bias applied to the charge conductor, $V_1-V_2$, but also the voltage applied to the heat source, $V_1-V_3$, plays a role in the charge current. Its effect is twofold: {\it (i)} it serves as a gate voltage to the conductor. This is evident in the charge current for the symmetric  configuration, $I^{(sym)}$, that develops a Coulomb blockade stability diagram; {\it (ii)} when $|\Theta_{3n}|>kT_3$, the charge fluctuations of the gate dot are suppressed (see figure \ref{occup}). Thus, in asymmetric configurations where, for instance $\Gamma_{l1}<\Gamma_{l0}$, the current through the conductor will be reduced in the region $\Theta_{31}<0$ where dot $g$ is almost constantly occupied, see the panel labeled $I^{(asym)}$ in figure \ref{curr}. This effect, which involves a competition of fast and slow transport transitions, leading to negative differential conductance, is known as dynamical channel blockade ~\cite{cottet}. 

The dynamical blocking of the current has visible effects only in configurations with energy-dependent tunneling. In the optimal configuration, where electrons in $s$ need to exchange energy with those in $g$ in order to contribute to the current, its effect is maximal. There, $I^{(opt)}$ is finite only in the fluctuating gate region (around $\Theta_{3n}=0$), where the charge of the hot dot fluctuates and heat flows, as shown in figure  \ref{curr}. Far from this region the heat current vanishes, i.e. the two conductors can be tuned to become thermally isolated by means of the gate voltage $V_3$. 

On the other hand, it is also necessary that the charge of the conductor fluctuates in order for heat to be transferred from the hot source. For the symmetric configuration, the heat flow is therefore suppressed on the Coulomb-blockade plateaux, and shows peaks of $J<0$ (out of terminal 3) along the lines $\Theta_{l0}{=}0$ and $\Theta_{l1}{=}0$, with $l{=}1,2$, where the Coulomb-blockade borders cross the fluctuating gate region. For larger applied bias, Joule heating in the conductor dominates such that the heat flow is reversed: i.e. transport pumps heat against the temperature gradient. Note that the generation of Joule heat prevents this effect to be used for refrigeration of the charge conductor. 

In the optimal configuration, an electron that has tunneled into dot $s$ needs to absorb or relax an energy $E_C$ before it can be transferred to the opposite contact. The sign of the heat flow is hence defined by the sign of the charge current: heat is driven by charge transport. For low voltages, a current flows in the direction defined by the asymmetry, e.g. from left to right if $\Gamma_{11}=\Gamma_{20}=0$. Applying a bias opposite to it, a stall potential $q(V_2-V_1)=E_C\eta_c$ is reached where the charge current generated by heat conversion is compensated by voltage. There, $I^{(opt)}=J^{(opt)}=0$ and the engine works reversibly~\cite{detection} at Carnot efficiency, $\eta_c=1-T/T_3$~\cite{hotspots}. A larger bias reverses the flow of charge and heat, such that heat is transferred to the hot terminal. On the contrary, if the bias is applied in the direction of the generated current, $I^{(opt)}$ and therefore also $J^{(opt)}$ will increase in absolute value, as compared with non-optimal configurations. In this way, driving a current through the conductor can serve to increase the heat extraction rate from the hot source.  For asymmetric enough configurations, this effect coexists with the heat pumping discussed above for the symmetric configuration, see $J^{(asym)}$ in figure \ref{curr}. 

\subsection{Noise and cross-correlations}

\begin{figure}[t]
\begin{center}
\includegraphics[width=0.9\linewidth,clip]{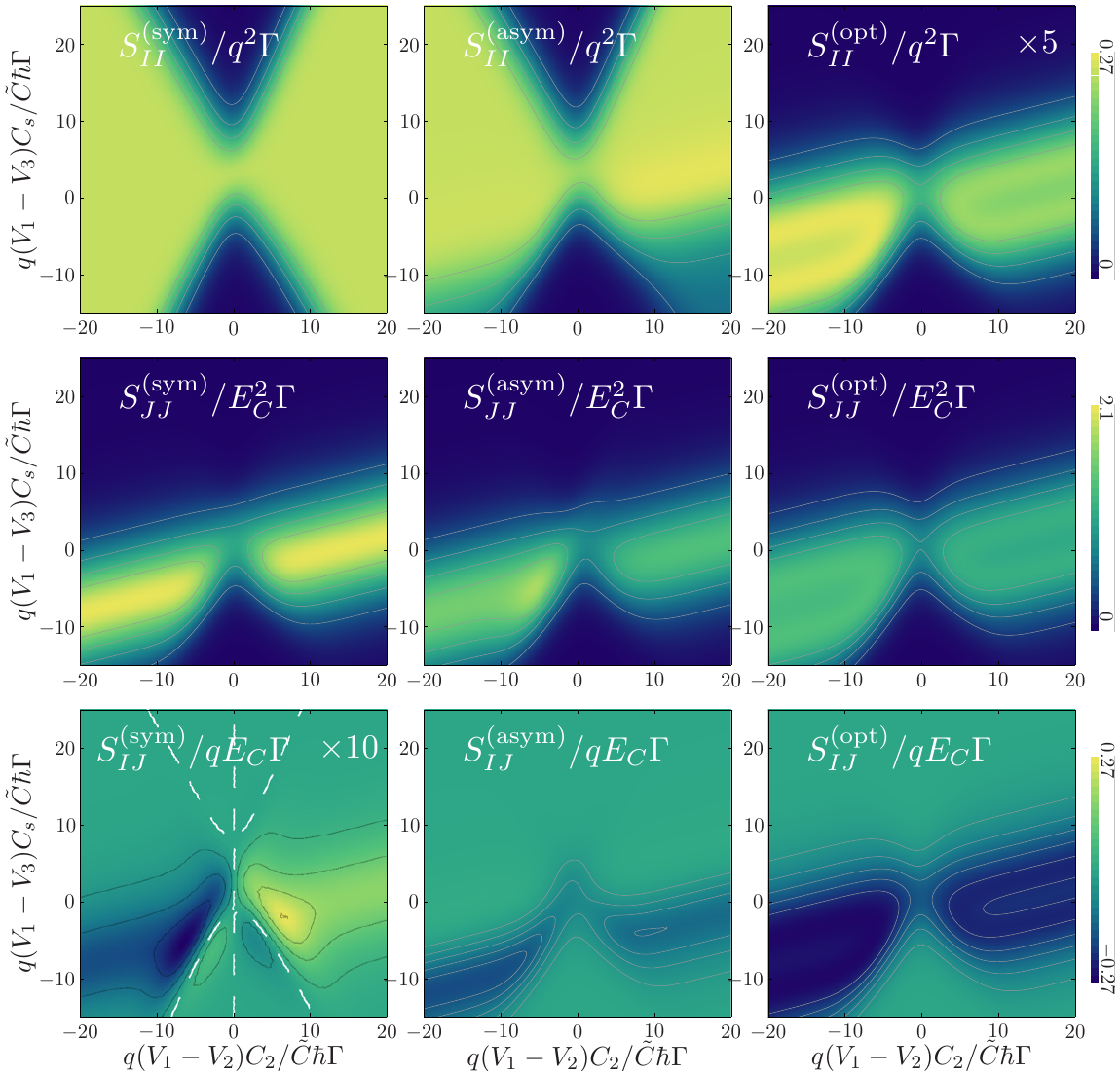}
\end{center}
\caption{\label{noise} Voltage dependence of the charge and heat noise, $S_{II}$ and $S_{JJ}$, and cross-correlations, $S_{IJ}$. The same tunneling configurations as in figure \ref{curr} are considered. Excess charge fluctuations are only noticeable out of the Coulomb blockade regions and constant for the symmetric configuration. 
The coupling to fluctuations in dot $g$ is visible in $S_{II}$ only for asymmetric configurations.
The cross-correlations change sign at the onset of charge transport, marked by white dashed lines, where Joule heating starts to dominate. In the optimal configuration all the correlation functions are proportional to each other.}
\end{figure}

Let us now discuss the current-current correlations for the different configurations. In the absence of noise-induced transport (symmetric configuration), the hot gate does not leave any clear feature in the charge noise: $S_{II}^{(sym)}$ presents similar characteristics as a (two-terminal) single quantum dot, i.e. a plateau dominated by shot noise (for large bias), and noise suppression in the Coulomb blockade region, cf. figure \ref{noise}. Obviously, the heat noise is also suppressed where $J$ vanishes. More interestingly, the charge-heat cross-correlations, $S_{IJ}^{(sym)}$, are maximal and change sign around the border of the Coulomb blockade region, related to the opening of conduction channels in $s$ and with the change of sign of $J^{(sym)}$ (see figure \ref{curr}), respectively.

The asymmetry in the energy-dependent couplings introduces features in the charge noise that depend on $\langle n_g\rangle$. As for the current, and depending on the asymmetry, the charge noise is reduced far from the fluctuating gate region, see figure \ref{noise}. 
On the other hand, the noise is maximal along $\Theta_{30}=0$ for $\Delta V>0$. It is a consequence of the dynamical channel blockade effect where a competition of slow and fast channels enhances the noise signal.  In the optimal configuration, excess noise and cross-correlations only appear in the region where both $n_s$ and $n_g$ fluctuate. Remarkably, in this configuration, $I^{(opt)}/q=J^{(opt)}/E_C$ and all the charge and heat correlations are proportional to each other:
\be
q^{-2}S_{II}^{(opt)}=E_C^{-2}S_{JJ}^{(opt)}=-(qE_C)^{-1}S_{IJ}^{(opt)}, 
\ee
consistently with our previous finding that $r^2=1$, see section \ref{sec:model}. 

\subsection{Fano factors}

\begin{figure}[t]
\begin{center}
\includegraphics[width=0.9\linewidth,clip]{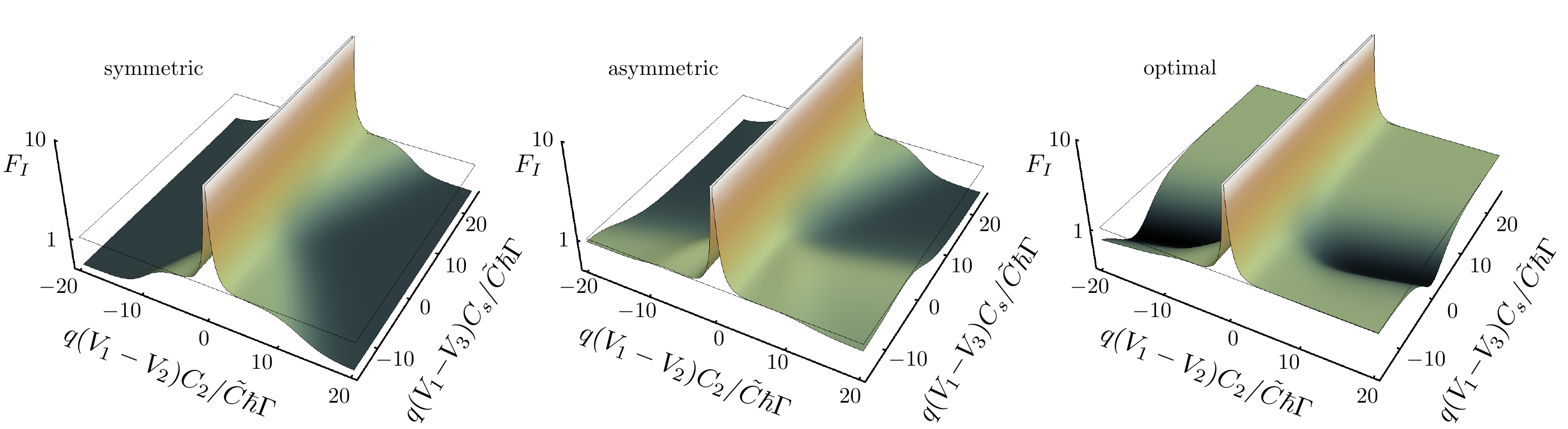}
\includegraphics[width=0.9\linewidth,clip]{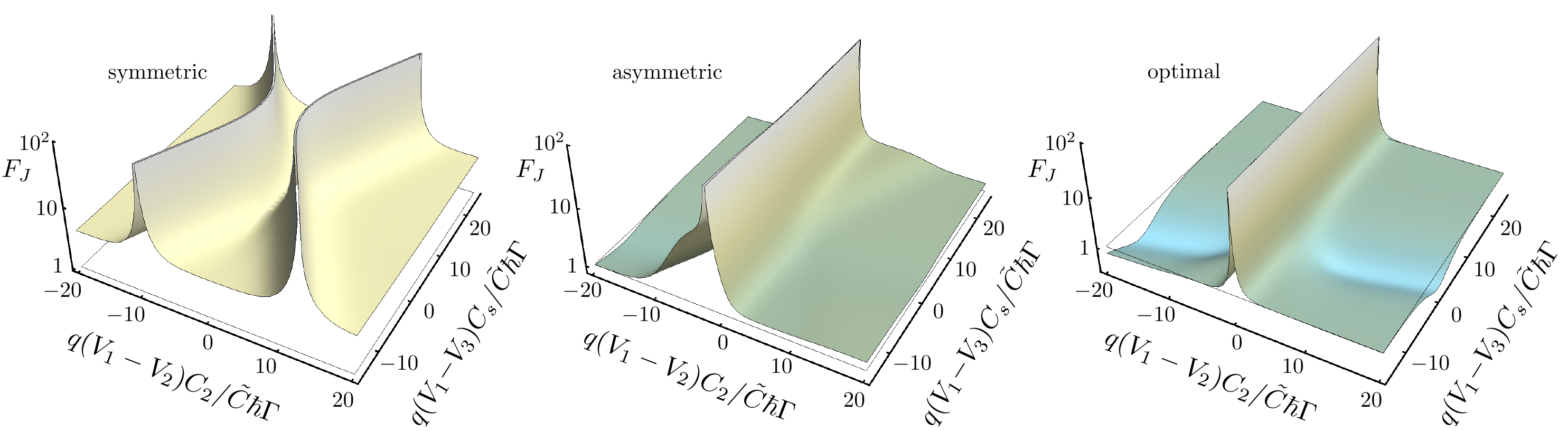}
\end{center}
\caption{\label{fano} Charge and heat Fano factors, $F_I$ and $F_J$ corresponding to the current and noise represented in figures \ref{curr} and \ref{noise}. Divergences appear when the currents vanish. }
\end{figure}

Additional information is obtained by looking at the noise to signal ratios: the Fano factors $F_I$ and $F_J$. They are depicted in figure \ref{fano} for the different tunneling configurations. Again, the charge statistics for the symmetric configuration is not changed by the presence of the hot terminal. Similar to a single-level two-terminal quantum dot, the Fano factor is close to Poissonian inside the Coulomb blockade region and $F_I^{(sym)}=1/2$ in the presence of transport~\cite{hershfield,korotkov}. We emphasize that higher order (co)tunneling processes which might affect the blocked region and may cause super-Poissonian noise~\cite{eugene,thielmann} are neglected here. At zero applied voltage, $I^{(sym)}=0$ and hence the Fano factor diverges. Similar divergences are observed in the heat Fano factor along the lines where Joule heating in the conductor compensates the flux coming from the heat source. The heat transfer depends on the characteristics of charge fluctuations in $s$, which is manifested in the Fano factor. Whereas both the heat current and noise vanish far from the condition $\Theta_{3n}=0$, $F_J^{(sym)}$ distinguishes the Coulomb blockade region and the large bias regime dominated by shot noise. In the large bias regime, $q(V_1-V_2)\gg kT$, it is given by: 
\be
F_J^{(sym)}=\mathrm{coth}\frac{E_C}{2kT_3}-c\mathrm{\ sech}\frac{\Theta_{30}}{2kT_3}\mathrm{\ sech}\frac{\Theta_{31}}{2kT_3}\sinh\frac{E_C}{2kT_3},
\ee
with $c=1/4+\Gamma\Gamma_g/[2(\Gamma_g+2\Gamma)^2]$ and $\Gamma_g=\Gamma_{30}=\Gamma_{31}$. It has a minimum at $\Theta_{30}=0$, where it is simply $F_J^{(sym)}=\mathrm{coth}(E_C\beta_3/2)-c\tanh (E_C\beta_3/2)$. Note that $F_J^{(sym)}$ is not necessarily greater than 1. For the regime that we are interested in, with $E_C\sim kT_3$, it is, however, super-Poissonian at any voltage.

Asymmetric configurations, where heat to charge conversion occurs, present peculiar features both in the charge and heat Fano factors. In the previous section, we discussed an increase of charge noise $S_{II}^{(asym)}$ in the regions where heat is most effectively transferred (along $\Theta_{3n}=0$) in terms of a dynamical blocking of the charge current due to the stabilization of the charge state in the hot dot.
Such an effect is known to lead to super-Poissonian noise~\cite{cottet,belzig,barthold,dqd} and positive cross-correlations~\cite{mcclure} in capacitively coupled conductors. When $\Gamma_{l1}\ll\Gamma_{l0}$, current only flows when dot $g$ is empty, i.e. $n_g$  acts as a current switch. If the switch rate is slow (for low $T_3$ and $\Gamma_{3n}$), the charge noise becomes super-Poissonian~\cite{andrew} (not shown). Consistent with telegraph noise, $F_I^{(asym)}$ diverges as $1/\Gamma_{3n}$. In the configuration relevant for heat engines, $T_3>T$ and $\Gamma_{3n}\sim\Gamma$, we find $F_I^{(asym)}\approx1$, cf. figure \ref{fano}. 
As a counterpart, the heat Fano factor becomes asymmetric and adopts some electron-like structure in the form of plateaux in the Coulomb blockade regions.

\begin{figure}[t]
\begin{center}
\includegraphics[width=0.9\linewidth,clip]{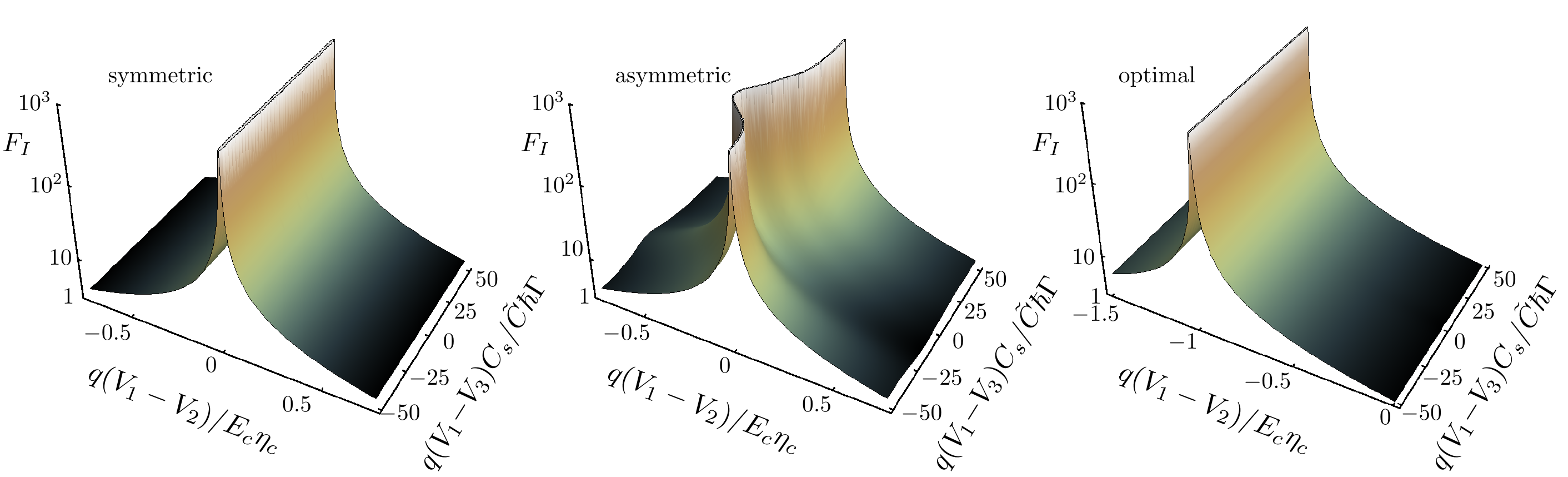}
\end{center}
\caption{\label{chargefanodet} Charge Fano factor for the different tunneling configurations close to zero bias, $V_1-V_2$. In the presence of energy dependent tunneling (asymmetric and optimal configurations), the Fano factor diverges at the stall potential where the applied voltage compensates the current generated by heat conversion. In the optimal case, it is given by $q(V_1-V_2)=-E_C\eta_c$. Here, $T_3=4T$.}
\end{figure}

An interesting feature is that, due to the noise-induced current, the charge Fano factor is finite at zero applied voltage. The divergence of $F_I$ is shifted to a finite stall voltage applied against the generated current such that $I^{(asym)}=0$. This non-equilibrium state (without charge current) is maintained by a finite heat current.
The point at which $F_I$ diverges can thus be considered as a measure of the non-linear thermovoltage, which
depends on the gate voltage, the asymmetry and the temperature gradient, as shown in figure \ref{chargefanodet}.

The optimal configuration is quite different. As we discussed above, transport is finite only in the region where the charge of the hot dot fluctuates, and charge and heat statistics become identical, with $F_I^{(opt)}=F_J^{(opt)}$. We recall that in this case there is only one tunneling sequence that contributes to charge and heat transport (the one represented in figure \ref{sys}). As a consequence, rather than a competition of different channels --which led to an increased charge noise in the asymmetric configuration-- they cooperate  and far from the divergence $F_I\le1$. Importantly, in the optimal configuration also the heat noise becomes sub-Poissonian, cf. figure \ref{fano}.  In this configuration, the divergence of the Fano factor is shifted to a constant stall voltage, $q(V_2-V_1)=E_C\eta_c$, where both $I^{(opt)}$ and $J^{(opt)}$ vanish, cf. figure \ref{chargefanodet}. Note that despite the non-equilibrium situation, no average current flows and entropy production is zero~\cite{detection}. Therefore, in the optimal case, at the stall voltage, the non-equilibrium state is maintained solely by fluctuations in the charge and heat current. 

We emphasize that in a Coulomb-blockade device like the one discussed here, charge fluctuations are antibunched even when noise is super-Poissonian \cite{clive}. This applies also to the heat transfer, which requires a sequence of charge tunneling events.

\section{Conclusions}

We have investigated the noise and cross-correlations of charge and heat currents in a quantum-dot energy harvester which can be measured by a charge detector. We demonstrated that finite equilibrium cross-correlations are related to the charge current generated by conversion of heat flow between conductors held at different temperatures~\cite{hotspots,detection,cavities} and rely on the energy-dependent asymmetry of the tunnel couplings. 
The properties of charge and heat fluctuations are mutually correlated also in the nonlinear regime. Charge noise increases in regions with heat fluctuations, and heat noise adopts charge-like statistics. We found different configurations where the heat source acts as a charge-current switch, and where the heat flux can be reversed by voltage such that it flows into the hot region. Depending on the asymmetry, transitions between sub- and super-Poissonian noise were predicted both for charge and heat currents. An optimal configuration where every quantum of transferred heat involves the transport of an electron across the conductor was identified and shown to correspond to a maximal charge-heat cross-correlation. 
We demonstrated that noise-induced transport manifests itself in shifting the divergence of the charge Fano factor to finite bias voltages. Also, the heat Fano factor is sensitive to the statistics of the fluctuations in the conduction dot, as well as to the energy dependence of its barriers.

\ack
We thank F. Hartmann, L. Worschech, H. Thierschmann, H. Buhmann and L. Molenkamp for discussions.
We acknowledge financial support from the Spanish MICINN Juan de la  
Cierva program and MAT2011-24331, the ITN Grant No. 234970 (EU), the Swiss NSF, the European STREP project Nanopower and the US NSF Grant No. DMR-0844899. 

\appendix
\section*{Appendix}
The dot internal potentials, $\phi_\alpha$, are given by the Poisson equations for the charge $Q_\alpha$:
\beann
Q_s=\sum_{l=1}^2C_l(\phi_s-V_l)+C(\phi_s-\phi_g)\nonumber\\
Q_g=C_3(\phi_g-V_3)+C(\phi_g-\phi_s),
\eeann
where $V_l$ is the voltage applied to terminal $l$. Note that the hot source also acts as a gate to the charge conductor. The electrostatic energy of each charge distribution, $U(Q_s,Q_g)$, is obtained by integrating: $U(Q_s,Q_g)=\sum_\alpha\int_0^{Q_\alpha}dQ_{\alpha}'\phi_\alpha$.
It defines the charging energies $U_{s,n}=U(1,n)-U(0,n)$ and $U_{g,n}=U(n,1)-U(n,0)$ given by:
\bea
U_{s,0}=\frac{q}{C\tilde C}\left(\frac{q}{2}C_{\Sigma g}+C_{\Sigma g}\sum_{l=1}^2C_lV_l+CC_3V_3\right)\nonumber\\
U_{g,0}=\frac{q}{C\tilde C}\left(\frac{q}{2}C_{\Sigma s}+C_{\Sigma s}C_3V_3+C\sum_{l=1}^2C_lV_l\right)\\
U_{\alpha,1}=U_{\alpha,0}+E_C,\nonumber
\eea
where $C_{\Sigma s}=C_1+C_2+C$ and $C_{\Sigma g}=C_3+C$ are the total geometric capacitance of each dot, and $\tilde C=(C_{\Sigma s}C_{\Sigma g}-C^2)/C$.

\end{document}